\documentclass[final,5p,twocolumn,times,11pt]{elsarticle}
\usepackage{graphicx}  
\usepackage{color}     
\usepackage{epsfig}    
\usepackage{float}    
\usepackage{bm}        
\usepackage{amssymb}   

\advance\voffset -0.2in
\biboptions{sort&compress}
\newcommand{\beq}{\begin{equation}}
\newcommand{\eeq}{\end{equation}}
\newcommand{\beqar}{\begin{eqnarray}}
\newcommand{\eeqar}{\end{eqnarray}}
\newcommand{\ds}{\displaystyle}
\begin{document}

\begin{frontmatter}

\title{ Is different $\Lambda$ and $\bar{\Lambda}$ polarization 
        caused by different spatio-temporal freeze-out picture?}

\author{O.~Vitiuk$^{1,2}$}

\author{L.V.~Bravina\corref{cor1}$^1$}
\ead{larissa.bravina@fys.uio.no}
\cortext[cor1]{Corresponding author}

\author{E.E.~Zabrodin$^{1,3}$}

\address{$^1$ Department of Physics, University of Oslo, PB 1048 
              Blindern, N-0316 Oslo, Norway \\
$^2$ Taras Shevchenko National University of Kyiv, UA-01033 Kyiv, 
     Ukraine \\
$^3$ Skobeltzyn Institute for Nuclear Physics,
     Moscow State University, RU-119899 Moscow, Russia }

\date{\today}

\begin{abstract}
Thermal vorticity in non-central Au+Au collisions at energies $7.7 
\leq \sqrt{s} \leq 62.4$~GeV is calculated within the UrQMD transport 
model. Tracing the $\Lambda$ and $\bar{\Lambda}$ hyperons back to their
last interaction point we were able to obtain the temperature and the 
chemical potentials at the time of emission by fitting the extracted 
bulk characteristics of hot and dense medium to statistical model of 
ideal hadron gas. Then the polarization of both hyperons was calculated.
The polarization of $\Lambda$ and $\bar{\Lambda}$ increases with 
decreasing energy of nuclear collisions. The stronger polarization of 
$\bar{\Lambda}$ is explained by the different space-time distributions 
of $\Lambda$ and $\bar{\Lambda}$ and by different freeze-out conditions
of both hyperons. \\ 
\small{{\bf PACS:} 25.75.-q, 24.10.Lx, 24.10.Pa } \\
\small{{\bf Keywords:} relativistic heavy-ion collisions, vorticity,
polarization, transport models }\\
\end{abstract}



\end{frontmatter}

\section{Introduction}
\label{sec1}

Experiments with heavy-ion collisions at relativistic and 
ultrarelativistic energies aim to study properties of very hot and dense
nuclear matter, most likely, quark-gluon plasma (QGP) \cite{qm18}. It is 
generally believed that in the nuclear matter phase diagram the curve, 
corresponding to first-order phase transition between the QGP and
hadronic matter, ends up (with rising bombarding energy) in a
tricritical point, where the transition becomes of second order. Above
this energy one is dealing with the crossover type of the phase 
transition. The latter corresponds to conditions emerging in heavy-ion
collisions at top RHIC ($\sqrt{s} = 200$~GeV) and LHC ($\sqrt{s} = 
2.76$~TeV and 5.02~TeV) energies. Searching for the tricritical point is 
one of the goals of Beam Energy Scan (BES) program at RHIC, studies of 
Pb+Pb collisions at SPS CERN, and future experiments on coming soon
facilities NICA at JINR and FAIR at GSI.

Non-central heavy-ion collisions generate enormous orbital angular 
momenta of order up to $10^5 \hbar$. Thus, the created QGP should possess
extremely high vorticity, that can be probed by global polarization of
hyperons, such as $\Lambda$ or $\bar{\Lambda}$. The idea connecting 
vorticity of QGP fluid with hyperon polarization was put forward in
\cite{LiWa_05}. Nowadays it becomes a very popular branch of
researches in heavy-ion collisions, see, e.g., 
\cite{BPR_08,RST_10,HHW_11,Gao_12,BCZG_13,BCW_13,Bec_15,PPQW_16,Li_17,
Bec_17,ST_17,BGST_17,XWC_17,KB_17,IS_16,IS_17,KTV_18,IS_18,CKW_19,ITS_19,
ITS_2_19} and
references therein. Meantime, STAR Collaboration has observed a very
peculiar behavior in polarization of $\Lambda$ and $\bar{\Lambda}$ in
semi-peripheral gold-gold collisions at center-of-mass energies between
7.7~GeV and 200~GeV. It is well known that polarization of both, 
$\Lambda$ and $\bar{\Lambda}$, at $\sqrt{s} = 200$~GeV is consistent 
with zero \cite{star_0}. Recently, similar result for the polarization 
of both hyperons in Pb+Pb collisions at $\sqrt{s} = 2.76$~TeV and 
5.02~TeV was reported by ALICE Collaboration \cite{alice_1}. However,
the polarization steadily increases to 2\% with reducing c.m. energy to 
$\sqrt{s} = 7.7$~GeV. Meanwhile, polarization of $\bar{\Lambda}$ at 
$\sqrt{s} > 7.7$~GeV is a bit larger although consistent within the 
error bars with that of $\Lambda$. But at $\sqrt{s} = 7.7$~GeV it 
suddenly rises up to $(8.3 \pm 3)\%$ \cite{star_1}. In contrast, in 
kinetic and hydrodynamic models polarizations of $\Lambda$ and 
$\bar{\Lambda}$ are essentially the same 
\cite{XWC_17,KB_17,Li_17,IS_18,ITS_19,ITS_2_19}. To explain the 
difference in polarizations of both hyperons in heavy-ion collisions at 
intermediate and low energies the anomalous mechanism related to axial 
vortical effect has been invoked in \cite{ST_17,BGST_17}. Another 
possibility discussed in \cite{CKW_19} is the interaction between the 
spins of (anti)hyperons and the vorticity of the baryon current.

In the present paper we argue that the difference in $\Lambda$ and 
$\bar{\Lambda}$ polarizations can originate from the different 
space-time distributions and different freeze-out conditions of both 
hyperons. The UrQMD model \cite{urqmd_1,urqmd_2} is employed to
investigate polarization of $\Lambda$ and $\bar{\Lambda}$ hyperons in
non-central Au+Au collisions in the energy range $7.7 \leq \sqrt{s} 
\leq 62.4$~GeV. The model permits us to trace the history of each hadron
back to the last inelastic (chemical freeze-out) or elastic (thermal 
freeze-out) collision. After that one can investigate the conditions of 
local equilibrium in a given area of the phase space and extract 
temperature $T$, baryon chemical potential $\mu_B$, and strangeness
chemical potential $\mu_S$. Details of this procedure are discussed in 
Sec.~\ref{sec2}. Section~\ref{sec3} describes the model used to 
determine the polarization of $\Lambda$ and $\bar{\Lambda}$. Results of
our study are presented in Sec.~\ref{sec4}. In particular, we show that
differences in $\Lambda$ and $\bar{\Lambda}$ polarization can be linked 
to different space-time freeze-out conditions for both types of hyperons, 
because various areas of the expanding fireball possess different 
vorticities at different times. Finally, conclusions are drawn in 
Sec.~\ref{concl}.

\section{UrQMD and statistical model of ideal hadron gas}
\label{sec2}

Relaxation of hot and dense nuclear matter produced in relativistic
heavy-ion collisions to local chemical and thermal equilibrium within
in microscopic transport models was studied in, e.g., 
\cite{plb_98,jpg_99,prc_99,prc_01,prc_08,OP_16}. The
method suggested there to determine temperature and chemical potentials 
of hot and dense medium is as follows. The whole volume of the collision 
is subdivide into relatively small cells. After the generation of 
appropriately high number of collisions at the same c.m. energy and with
the same centrality, one can extract energy density $\varepsilon$, net 
baryon density $\rho_B$, and net strangeness density $\rho_S$. In 
equilibrium, the properties of mixture of hadron species are determined 
by the set of distribution functions (in system of natural units of $c = 
\hbar = k_B = 1$)
\beq \ds
  f(p,m_i) = \left[ \exp{\left( \frac{\epsilon_i - \mu_i}{T} \right)} 
             \pm 1 \right]^{-1}
\label{eq1}
\eeq
Here $p$ is the momentum, $m$ is the mass, $\epsilon_i = \sqrt{p^2 + 
m_i^2}$ is the energy, and $\mu_i$ is the chemical potential of the
hadron specie $i$, respectively. The sign $-$ stands for bosons and the
$+$ is for fermions. We will consider baryon and strangeness chemical 
potentials and disregard electric chemical potential, which is usually
much smaller than the first two ones. Therefore, the total chemical
potential of the $i$-th hadron with the baryon charge $B$ and strange 
charge $S$ reads
\beq \ds
  \mu_i = B_i \mu_{\rm B} + S_i \mu_{\rm S}
\label{eq2}
\eeq
The first and the second moments of the distribution function provide us
the particle number density $n_i$ and the energy density $\varepsilon_i$,
respectively
\beqar \ds
  n_i &=& \frac{g_i}{(2\pi)^3} \int f(p,m_i) d^3p \\
\label{eq3}
  \varepsilon_i &=& \frac{g_i}{(2\pi)^3} \int \epsilon_i f(p,m_i) d^3p
\label{eq4}
\eeqar
with $g_i$ being the spin-isospin degeneracy factor. In order to obtain 
$T$, $\mu_{\rm B}$ and $\mu_{\rm S}$ one has to insert the extracted 
microscopic parameters 
$\{\varepsilon^{mic}, \rho_{\rm B}^{mic}, \rho_{\rm S}^{mic}\}$ into the
system of nonlinear equations
\beqar \ds
\label{eq5}
\rho_{\rm B}^{mic} &=& \sum \limits_{i} B_i\, n_i (T, \mu_{\rm B}, 
                       \mu_{\rm S})\ , \\ 
\label{eq6}
\rho_{\rm S}^{mic} &=& \sum \limits_{i} S_i\, n_i (T, \mu_{\rm B}, 
                       \mu_{\rm S})\ , \\
\label{eq7}
\varepsilon^{mic} &=& \sum_i \varepsilon_i (T, \mu_{\rm B}, 
                       \mu_{\rm S})\ ,
\eeqar
Note that the set of hadron species employed in the SM must be identical 
to that of the microscopic model. Comparison of hadron yields and energy 
spectra, given by the microscopic model, to the SM spectra in the 
central area of heavy-ion collisions has revealed that hot and dense 
nuclear matter needed about $6 - 8$~fm/$c$ to reach the vicinity of 
chemical and local equilibrium.  The extracted temperature will be used 
for calculation of vorticity of nuclear matter. This issue is discussed
in Sec.~\ref{sec3}.
    
\section{Calculation of vorticity and $\Lambda, \bar{\Lambda}$
polarization}
\label{sec3}

There are several approaches to the vorticity problem in heavy-ion 
collisions. The most popular is the treatment based on the assumption of
local thermal equilibrium at the system freeze-out \cite{BCW_13}. 
Another mechanism is the axial charge separation due to the chiral 
vortical effect \cite{ST_17}. In the present paper we will follow 
Refs.~\cite{BCW_13,XWC_17}. If the system is in local equilibrium and
the concentration of both, $\Lambda$ and $\bar{\Lambda}$, is very small,
their ensemble averaged spin 4-vector at space-time point $x$ reads
\beq \ds
  S^{\mu}(x,p)=-\frac{1}{8m}\epsilon^{\mu\nu\rho\sigma}p_{\nu}\varpi_{
  \rho\sigma}(x) \ ,
\label{eq:spin_relativistic}
\eeq 
containing the hyperon mass $m$, antisymmetric tensor 
$\epsilon^{\mu\nu\rho\sigma}$, and the thermal vorticity tensor
\beq \ds
  \varpi_{\mu\nu}=\frac{1}{2}\left(\partial_{\nu}\beta_{\mu}-
  \partial_{\mu}\beta_{\nu}\right)
\label{eq:thermal_vorticity}
\eeq
Here $\beta^{\mu}=u^{\mu}/T$ is the inverse-temperature four-velocity, 
$u^{\mu}$ is hydrodynamic four-velocity and $T$ is a proper temperature,
respectively. Decomposing the thermal vorticity into the space-time
components,
\beqar \ds
  \boldsymbol{\varpi}_{T} &=& (\varpi_{0x},\varpi_{0y},\varpi_{0z}) =
  \frac{1}{2}\left[\nabla\left(\frac{\gamma}{T}\right)+\partial_{t}
  \left(\frac{\gamma\mathbf{v}}{T}\right)\right] \ , \\
\label{eq:vorticity_time}
  \boldsymbol{\varpi}_{S} &=& (\varpi_{yz},\varpi_{zx},\varpi_{xy}) =
  \frac{1}{2}\nabla\times\left(\frac{\gamma\mathbf{v}}{T}\right) \ ,
\label{eq:vorticity_space} 
\eeqar
one gets for the spin vector
\beqar \ds
  S^{0}(x,p) &=& \frac{1}{4m}\mathbf{p}\cdot\boldsymbol{\varpi}_{S}\ ,\\
\label{eq:spin_zero_component}
  \mathbf{S}(x,p) &=& \frac{1}{4m}\left(E_{p}\boldsymbol{\varpi}_{S} +
  \mathbf{p}\times\boldsymbol{\varpi}_{T}\right)\ ,
\label{eq:spin_vector_components}
\eeqar
where $E_p = \sqrt{\mathbf{p}^2 + m^2}$ is the energy, and $\mathbf{p}$
is the momentum of $\Lambda$.

The spin vector of $\Lambda$ hyperon measured in the STAR experiment in
the local rest frame of $\Lambda$, i.e. $S^{\ast\mu} = (0,
\mathbf{S}^\ast)$, is related to that in the center-of-mass frame of 
Au+Au collisions by a Lorentz boost
\beq \ds
  \mathbf{S}^\ast(x,p)=\mathbf{S}-\frac{\mathbf{p}\cdot\mathbf{S}}{E_{p}
  \left(m+E_{p}\right)}\mathbf{p}\ .
\label{eq:spin_rest_frame}
\eeq
In the transport model calculations we have to average vector 
$\mathbf{S}^\ast$ over all $\Lambda$'s emitted from the expanding 
fireball
\beq \ds
  \left\langle \mathbf{S}^\ast\right\rangle = \frac{1}{N}\sum_{i=1}^{N}
  \mathbf{S}^\ast(x_{i},p_{i}) \ ,
\label{eq:spin_average}
\eeq
where $N$ is the total amount of Lambdas in all events. The global 
polarization of $\Lambda$ in the STAR experiment is the projection of 
averaged spin $\left\langle \mathbf{S}^\ast\right\rangle$ onto the 
direction of global angular momentum in non-central collisions 
\cite{Bec_15}
\beq \ds
  P = \frac{\left\langle \mathbf{S}^\ast\right\rangle \cdot\mathbf{J}}
  {|\left\langle \mathbf{S}^\ast\right\rangle| |\mathbf{J}|} \ ,
\label{eq:polarization}
\eeq
with $\mathbf{J}$ being the global orbital angular momentum.

\section{Vorticity of nuclear matter and $\Lambda, \bar{\Lambda}$
         polarization}
\label{sec4}

We studied Au+Au collisions at c.m. energies corresponding to Beam 
Energy Scan program at RHIC, ranging from $\sqrt{s} = 7.7$ to 
62.4~GeV. At each energy one million collisions with the impact 
parameters $b = 6$~fm and $b = 9$~fm were generated. - The impact 
parameters were chosen to compare the results of $\Lambda$ and 
$\bar{\Lambda}$ polarization to the experimental data obtained at 
centrality $20\% \leq \sigma / \sigma_{geo} \leq 50\%$. The kinematic 
cuts imposed on the $\Lambda$ and $\bar{\Lambda}$ spectra match those 
of the STAR experiment \cite{star_1} for rapidity, $ |y| \leq 1$, and 
transverse momentum, $0.1 \leq p_T \leq 3$~GeV/$c$.

\begin{figure}
    \includegraphics[scale=0.50]{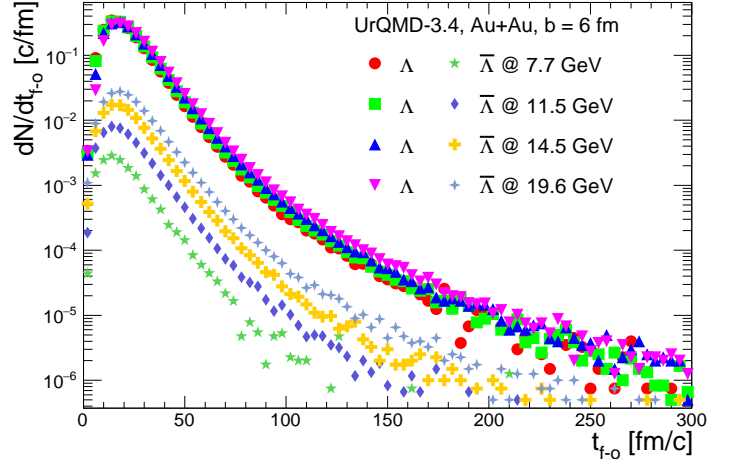}
\caption{(Color online) 
Emission functions for $\Lambda / \bar{\Lambda}$ hyperons 
in UrQMD calculations of Au+Au collisions with $b = 6$~fm at $\sqrt{s} = 
7.7$~GeV (circles/stars), 11.5~GeV (squares/diamonds), 14.5~GeV 
(triangles-up/big crosses), and 19.6~GeV (triangles-down/small crosses).}
\label{freeze}
\end{figure}

First, one has to study freeze-out conditions of both hyperons, because 
it is well-known that microscopic transport models demonstrate
non-uniform continuous freeze-out of hadrons 
\cite{Bass_FO_99,FO_plb_95,FO_prc_99} rather than the sharp one.
Emission functions $d N /dt$ for $\Lambda$ and $\bar{\Lambda}$ are 
shown in Fig~\ref{freeze}. 

Although the main amounts of both hyperons are
emitted within 10 - 25~fm/$c$, there is a continuous radiation of
$\Lambda$ and $\bar{\Lambda}$ up to very late stage. The mean values of 
the freeze-out times, listed in Table~\ref{tab1}, show that 
$\bar{\Lambda}$ hyperons are emitted in average about 1~fm/$c$ earlier
than $\Lambda$'s. This difference looks insignificantly small, however, 
the fireball rapidly expands and its temperature drops quickly. 
\begin{table}
\begin{center}
     \begin{tabular}{ l | l | l | l | l }
        \hline\hline
            $\sqrt{s}$ (GeV)&7.7&11.5&14.5&19.6 \\ \hline
            $\langle t_{\Lambda}^{FO} \rangle$ (fm/c) &21.3009 &   
                              21.9568 &    23.066  &   24.3462 \\
            $\langle t_{\bar{\Lambda}}^{FO} \rangle$ (fm/c) &19.7806 &   
                              21.0302 &    21.959  &   23.1288 \\ 
         \hline\hline
      \end{tabular}
\end{center}
\caption{Mean freeze-out time of $\Lambda$ and $\bar{\Lambda}$ hyperons 
in UrQMD calculations of Au+Au collisions with $b = 6$~fm at $\sqrt{s} = 
7.7-19.6$~GeV.}
\label{tab1}
\end{table}

\begin{figure}
  \resizebox{\linewidth}{!}{
       \includegraphics[width=\textwidth]{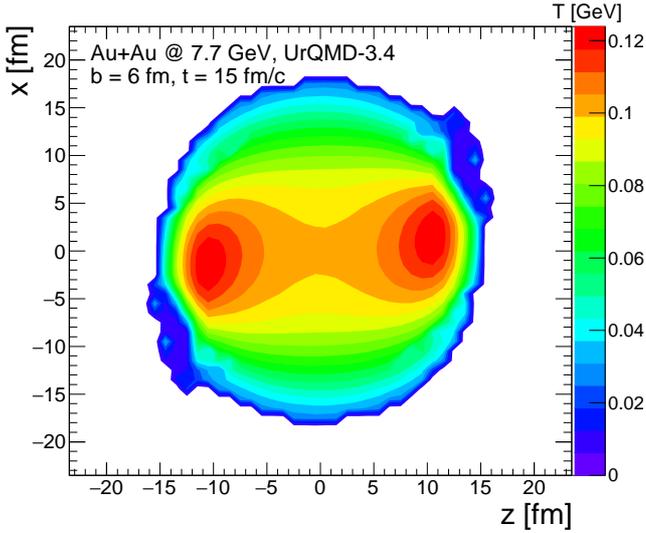}}
\caption{(Color online) 
Distribution of the proper temperature in the reaction plane 
in UrQMD calculations of Au+Au collisions with $b = 6$~fm at $\sqrt{s} = 
7.7$~GeV at $t = 15$~fm/$c$.}
\label{temp}
\end{figure}

We have to check, therefore, the temperatures of the areas from where 
the hyperons were emitted. To get the temperature map, the whole space 
was subdivided into cubic cell with volume $V = 1$~fm$^3$. Then, we 
calculated the total energy density $\varepsilon$, net baryon density
$\rho_{\rm B}$, and net strange density $\rho_{\rm S}$ as functions of 
time $t$ for each cell in its local rest frame. The time step is 
$\Delta t = 1$~fm/$c$. After that, the procedure described in 
Sec.~\ref{sec2} was employed to find the temperature and the chemical 
potentials in each cell. Figure~\ref{temp} displays the distribution of
temperature in the reaction plane of Au+Au collisions with $b = 6$~fm at
$\sqrt{s} = 7.7$~GeV at $t = 15$~fm/$c$ after beginning of the collision.

\begin{figure}[ht]
  \resizebox{\linewidth}{!}{
       \includegraphics[width=\textwidth]{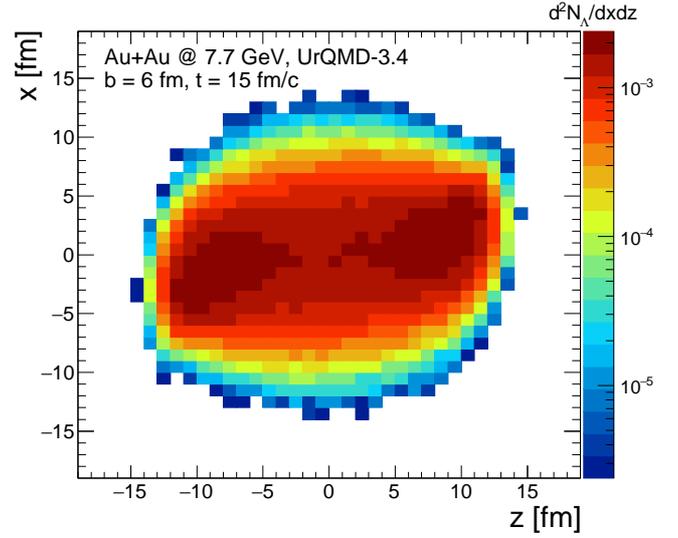}}
  \resizebox{\linewidth}{!}{
       \includegraphics[width=\textwidth]{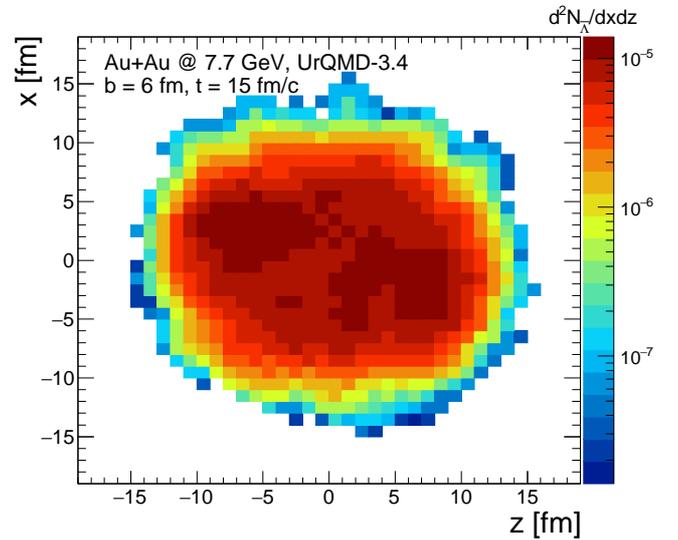}}
\caption{(Color online) 
Density distributions $d^2N / dx dz$ of $\Lambda$ (upper plot) and 
$\bar{\Lambda}$(bottom plot) in the reaction plane 
in UrQMD calculations of Au+Au collisions with $b = 6$~fm at $\sqrt{s} = 
7.7$~GeV at $t = 15$~fm/$c$.}
\label{density}
\end{figure}

One can see that the temperature is not uniformly distributed within the
whole volume. Here the spectator areas are the hottest parts of the 
fireball. The temperature gradually drops from the inner to outer zones.
Note again, that this is the proper temperature obtained after the
subtraction of collective velocity of each cell. Therefore, the 
temperatures of emitted hadrons depend both on the emission times and 
on the location of their freeze-out areas in space. The density 
distributions of both hyperons in the reaction plane $d^2N / dx dz$ 
depicted in Fig.~\ref{density} are quite different. Whereas maximum 
densities of $\Lambda$ are in the spectator's areas, $\bar{\Lambda}$ 
are concentrated mainly in the baryon-less zones. Thus, both $\Lambda$ 
and $\bar{\Lambda}$ are not only frozen, in average, at a bit different 
times. These hyperons are also emitted from different areas of space 
with different temperatures.

\begin{figure}
  \resizebox{\linewidth}{!}{
       \includegraphics[width=\textwidth]{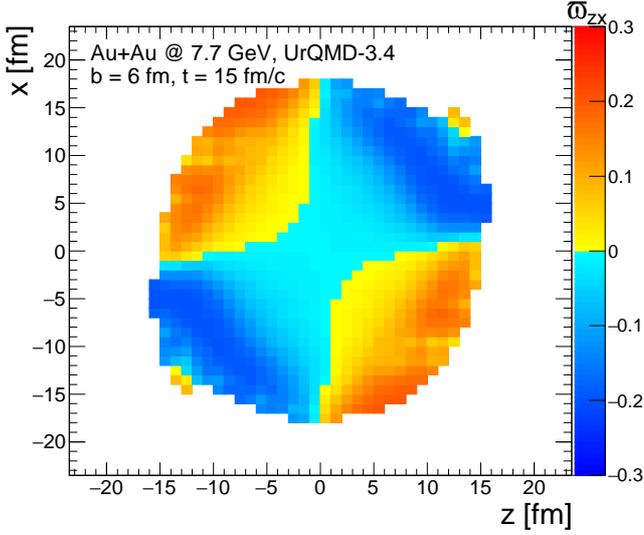}}
\caption{(Color online) 
Thermal vorticity component $\varpi_{zx}$ in the reaction plane 
in UrQMD calculations of Au+Au collisions with $b = 6$~fm at $\sqrt{s} = 
7.7$~GeV at $t = 15$~fm/$c$.}
\label{vort}
\end{figure}

We are ready now to study vorticity in the system. Of three vorticity
components in space, the reaction-plane component $\varpi_{zx}$ is the
most important for calculation of $\Lambda$ and $\bar{\Lambda}$
polarization because it is parallel to angular momentum of the system.
This vorticity is presented in Fig.~\ref{vort}; the calculations are 
also done for gold-gold collisions with $b = 6$~fm at $\sqrt{s} = 
7.7$~GeV and at $t = 15$~fm/$c$.

Although $\varpi_{zx}$ has a quadruple-like structure in $z-x$ plane, 
its first and third quadrants are connected by region with small negative 
vorticity. This connecting part becomes smaller and disappears with 
increasing collision energy, whereas at lower energies it becomes larger. 
The structure is stable in time, but the vorticity magnitude decreases 
due to system expansion. It means that in general the average value of 
$\varpi_{zx}$ is negative, thus resulting to positive total polarization. 
Also, it means that the global polarization of hyperons should decrease 
with time and with rising energy of the collision.
Our result is compatible with other transport model calculations 
\cite{Li_17} and with hydrodynamic calculations of thermal vorticity 
\cite{Bec_15,ITS_19}.
            
\begin{figure}[!htbp]
  \resizebox{\linewidth}{!}{
    \includegraphics[width=\textwidth]{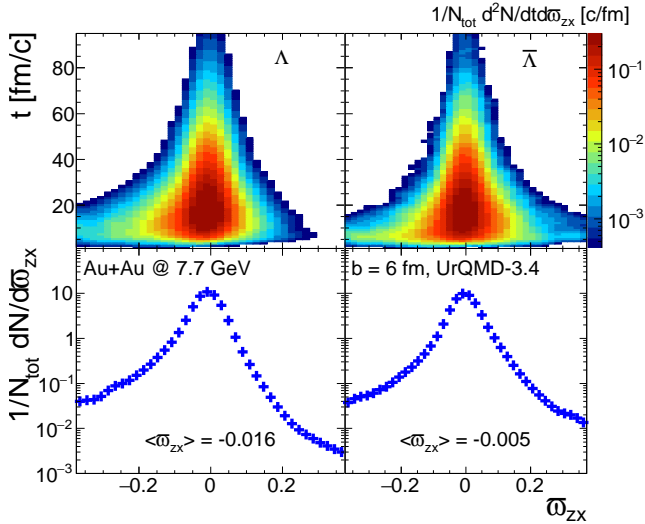}}
\caption{(Color online) 
Upper row: The emission function of $\Lambda$ (left) and 
$\bar{\Lambda}$ (right) as function of time and $\varpi_{zx}$ component 
of vorticity in UrQMD calculations of Au+Au collisions with $b = 6$~fm 
at $\sqrt{s} = 7.7$~GeV.
Bottom row: The distribution of $\Lambda$ (left) and $\bar{\Lambda}$ 
(right) over $\varpi_{zx}$ component of vorticity at the emission point.}
\label{emis7}
\end{figure}

\begin{figure}[ht]
  \resizebox{\linewidth}{!}{
    \includegraphics[width=\textwidth]{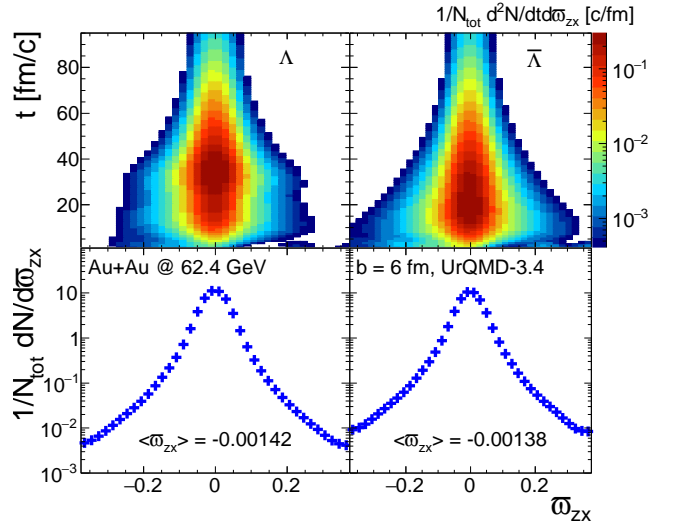}}
\caption{(Color online) 
The same as Fig.~\ref{emis7}, but for $\sqrt{s} = 62.4$~GeV.}
\label{emis19}
\end{figure}

Now we will analyze Lambda and antiLambda hyperons distribution over 
$\varpi_{zx}$ at freeze-out point and study their emission functions as 
functions of both time and $\varpi_{zx}$. These functions are displayed 
in Fig.~\ref{emis7} and Fig.~\ref{emis19} for UrQMD calculations of Au+Au 
non-central collisions with $b = 6$~fm at $\sqrt{s} = 7.7$~GeV and 
$62.4$~GeV, respectively.

\begin{figure}[!htbp]
  \resizebox{\linewidth}{!}{
    \includegraphics[width=\textwidth]{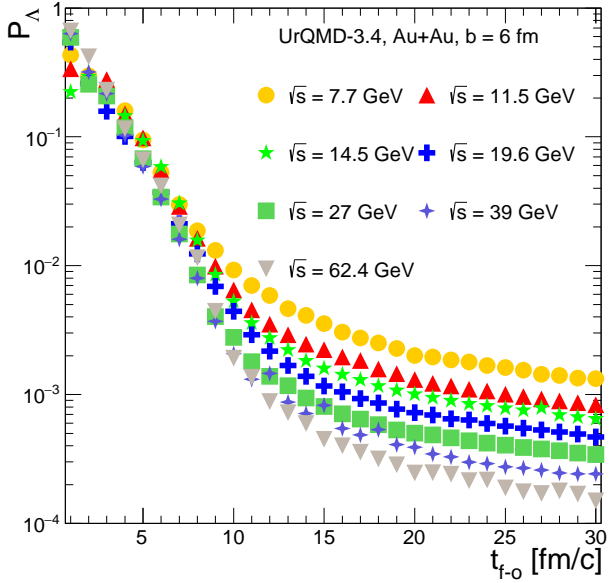}}
  \resizebox{\linewidth}{!}{
    \includegraphics[width=\textwidth]{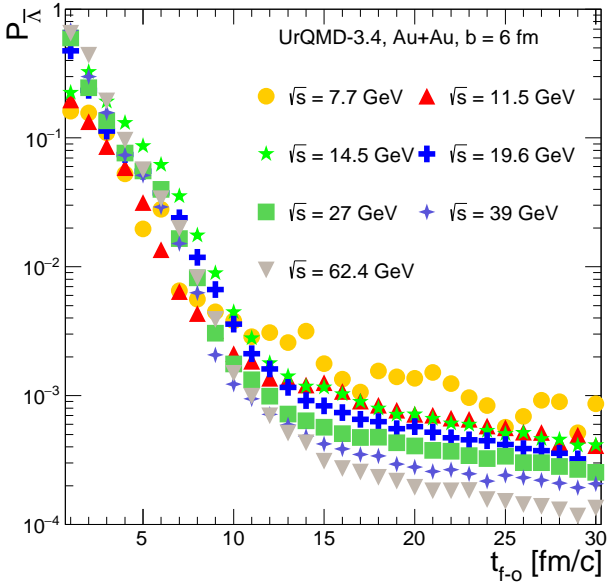}}
\caption{(Color online) 
Upper plot: Global $\Lambda$ polarization dependence on the emission 
time for $\sqrt{s}=7.7$~GeV (circles), 11.5~GeV (triangles up), 14.5~GeV 
(stars), 19.6~GeV (crosses), 27~GeV (squares), 39~GeV (asterisks) and
62.4~GeV (triangles down), respectively. Bottom plot: The same as the 
upper plot but for $\bar{\Lambda}$ polarization.}
\label{poltime}
\end{figure}

At $\sqrt{s} = 7.7$~GeV $\Lambda$ and $\bar{\Lambda}$ are mainly emitted 
from the regions with small negative vorticity, thus both hyperons have 
non-zero positive polarization. The distributions of $\Lambda$ and 
$\bar{\Lambda}$ have pronounced maxima shifted to negative values of 
$\varpi_{zx}$. For $\Lambda$ the maximum is located at 
$\varpi_{zx}^\Lambda \simeq -0.016$, whereas for $\bar{\Lambda}$ the 
maximum is shifted closer to zero, $\varpi_{zx}^{\bar{\Lambda}} \simeq 
-0.005$. At $\sqrt{s} = 62.4$~GeV, shown in Fig.~\ref{emis19}, both 
hyperon species have more symmetric and wide distributions. The 
positions of their maxima are at 
$\varpi_{zx}^{\Lambda/\bar{\Lambda}} \simeq -0.0014$. Emission functions 
of both $\Lambda$ and $\bar{\Lambda}$ show that the main bunch of 
$\Lambda$ and $\bar{\Lambda}$ is decoupled from the system between 
7~fm/$c$ and 25~fm/$c$. The distributions of the later emitted hyperons 
are more symmetric with respect to $\varpi_{zx}$ thus implying that the 
overall hyperon polarization is dominated by early emitted particles.

emission point, one can calculate the global polarization of both
hyperon species. Recall that polarization of each hyperon is calculated 
with Eq.(\ref{eq:spin_relativistic}) using the thermal vorticity at the 
space-time point of the hyperon production.
Global polarizations of $\Lambda$ and $\bar{\Lambda}$
as functions of the emission time are presented in Fig.~\ref{poltime}
for energies ranging from $\sqrt{s} = 7.7$~GeV to $\sqrt{s} = 62.4$~GeV.
Both polarizations quickly drop almost exponentially from 20-60\% at 
very early times to about 0.5\% at $t = 10$~fm/$c$. After that time the
slopes of the distributions become more acclivous. The higher the energy
of the collision, the steeper the slopes. The explanation of this effect 
is rather straightforward. At the very beginning all $\Lambda$ and 
$\bar{\Lambda}$ are formed within the lump of hot and dense matter with 
high polarization. As the fireball expands, the average vorticity 
$\varpi_{zx}$ and polarization in different areas of the fireball 
decrease rapidly. The larger global polarization at lower bombarding
energies originates from the larger production rates of $\Lambda$ and 
$\bar{\Lambda}$ in the negative-vorticity region because of the slower
expansion rate of the fireball.
 
\begin{figure}[ht]
  \resizebox{\linewidth}{!}{
   \includegraphics[width=0.95\textwidth]{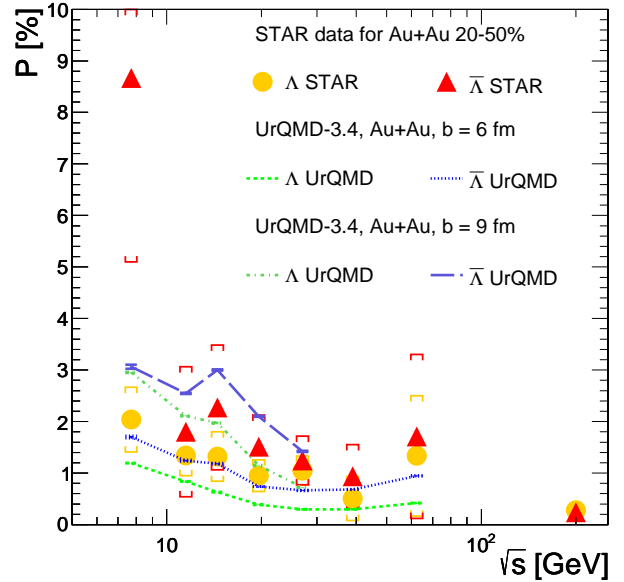}}
\caption{(Color online) 
Global $\Lambda$ (dashed line) and $\bar{\Lambda}$ (dotted line) 
polarization as function of $\sqrt{s}$ in UrQMD calculations of Au+Au 
collisions with $b = 6$~fm in comparison with the data from STAR 
experiment \cite{star_1,star_2} (circles and triangles, respectively).
Polarization of $\Lambda$ and $\bar{\Lambda}$ in Au+Au collisions with
$b = 9$~fm is shown by the dash-dotted and long-dashed lines, 
respectively. }
\label{polar}
\end{figure}

Using the thermal vorticity and momentum of each hyperon at its 
Energy dependence of the global polarization of $\Lambda$ and 
$\bar{\Lambda}$ in Au+Au collisions with $b = 6$~fm and $b = 9$~fm is 
displayed in Fig.~\ref{polar}. Here the results of the UrQMD 
calculations are confronted to the STAR data \cite{star_1,star_2} 
obtained for the centrality $20\% \leq \sigma/\sigma_{geo} \leq 50\%$. 
We took into account $\Lambda$ and $\bar{\Lambda}$ emitted before
$t = 30$~fm/$c$. For the collisions 
at energies $\sqrt{s} \geq 11.6$~GeV the UrQMD results are very close to 
the data. It is worth mentioning that the fixed impact parameter $b = 
6$~fm corresponds to centrality close to 20\%,
whereas $b = 9$~fm roughly corresponds to 45\% of centrality. For more 
peripheral collisions polarization of both hyperons increases. For all
energies except $\sqrt{s} = 7.7$~GeV the difference between the 
polarizations of $\Lambda$ and $\bar{\Lambda}$ in Au+Au collisions with 
$b = 9$~fm is at least not smaller than that in collisions with 
$b = 6$~fm. 
The increase of global polarization of 
both hyperons at lower energies is due to more abundant production of 
$\Lambda$ and $\bar{\Lambda}$ in the negative-vorticity region because
of the slow expansion rate. The difference between the global 
polarization of $\Lambda$ and $\bar{\Lambda}$, clearly seen in 
experimental data, is correctly reproduced in the model. It is 
explained by the difference in space-time distributions of $\Lambda$ and 
$\bar{\Lambda}$ and different freeze-out conditions of both hyperons 
with respect to the thermal vorticity field. 

Note that the polarization of $\Lambda$ hyperons produced in decays of
$\Sigma^\ast(1385)$ or $\Sigma^0(1192)$ was calculated similarly to the
polarization of primarily produced $\Lambda$. As shown, e.g., in 
\cite{Bec_17} and \cite{KB_17}, the decays of $\Sigma$ lead to 
suppression of the global polarization of $\Lambda$ by about 15-20\%. 
However, the $\Sigma$ hyperons in microscopic transport calculations 
possess the thermal vorticity of the cell they were produced or 
experienced the last interaction before the decay to $\Lambda + \pi$ for 
$\Sigma^\ast$ or $\Lambda + \gamma$ for $\Sigma^0$. This vorticity is 
larger than the thermal vorticity of the cell where the decay of 
$\Sigma$ takes place. We checked the interplay of both effects for Au+Au 
collisions with $b = 6$~fm at $\sqrt{s} = 7.7$~GeV. The overall 
correction to global $\Lambda$ polarization was found to be about 4\%, 
i.e., quite small. The detailed study of this interplay for other 
energies and impact parameters will be done in a forthcoming publication.

The model cannot match only the magnitude of $\bar{\Lambda}$
polarization measured at $\sqrt{s} = 7.7$~GeV. 
Here the polarization of both hyperons in collisions with $b = 9$~fm
increases to approximately 3\%, see Fig.~\ref{polar}. Note, however, 
that the yield of $\bar{\Lambda}$ in peripheral Au+Au collisions at 
this energy is very low. Thus, the relatively modest value of 
$\bar{\Lambda}$ polarization can be caused by unfortunate statistical 
fluctuation. Another plausible solution is to increase slightly the   
number of antilambdas emitted within first 10~fm/$c$.
The sudden rise of $\bar{\Lambda}$ polarization at low energies deserves 
further investigations.    

\section{Conclusions}
\label{concl}

We calculated thermal vorticity in Au+Au collisions with $b = 6$~fm at
beam energy scan energies $7.7 \leq \sqrt{s} \leq 62.4$~GeV within the 
UrQMD model. Statistical model of ideal hadron gas was used to extract 
the temperature of the space areas from where both $\Lambda$ and 
$\bar{\Lambda}$ were emitted. Quadruple structure of $\varpi_{zx}$ 
component of vorticity is obtained, and dependence of $\varpi_{zx}$ on
time and energy is studied.

It was found that freeze-out conditions of $\Lambda$ and $\bar{\Lambda}$ 
are different both in space and in time. This means, particularly, that 
the studied hyperons are emitted from parts of the fireball with 
different thermal vorticity.

Method for calculation of $\Lambda$ and $\bar{\Lambda}$ global 
polarization in transport model is developed.
Using this method $\Lambda $ and $\bar{\Lambda}$ global polarization is 
calculated in Au+Au collisions at energies $\sqrt{s}=7.7-62.4$~GeV and 
compared with the data from STAR experiment.
Polarization of $\Lambda$ and $\bar{\Lambda}$ decreases with increasing
collision energy in line with the experimental data.

It was found that within the thermal approach the global polarization 
in transport models is jointly determined by the space-time distribution 
of $\Lambda / \bar{\Lambda}$ and the thermal vorticity field. The larger 
global polarization at lower collision energies is due to larger 
production rate of the hyperons in the negative-vorticity region because 
of the slow expansion rate. This means that the magnitude of vorticity 
decreases slower than at higher collision energies.

For the first time, the difference between global polarization of 
$\Lambda$ and $\bar{\Lambda}$ is obtained within the thermal approach. 
This difference is naturally explained by the difference in space-time 
distributions of $\Lambda$ and $\bar{\Lambda}$ and different freeze-out 
with respect to the thermal vorticity field.

{\bf Acknowledgments.}
The authors are grateful to L.~Csernai, Yu.~Ivanov, A.~Sorin and
O.~Teryaev for fruitful discussions and valuable comments. 
The work of L.B. and E.Z. was supported by Russian Foundation for Basic
Research (RFBR) under Grants No. 18-02-40084 and No. 18-02-40085,
and by the Norwegian Research Council (NFR) under Grant No. 255253/F50 -
``CERN Heavy Ion Theory". O.V. acknowledges the financial support of the 
Norwegian Centre for International Cooperation in Education (SIU) under 
Grant ``CPEA-LT-2016/10094 - From Strong Interacting Matter to Dark 
Matter". All computer calculations were made at Abel (UiO, Oslo) and 
Govorun (JINR, Dubna) computer cluster facilities. 


\end{document}